\documentclass[final,5p,times,twocolumn]{elsarticle}
\usepackage{hyperref}
\usepackage{multirow}
\usepackage{graphicx}
\usepackage{amsmath}
\usepackage{textcomp}
\usepackage{txfonts}
\usepackage{lineno}
\modulolinenumbers[5]
\journal{Nuclear Instruments and Methods in Physics Research A}
\begin{document}

\begin{frontmatter}

\title{A 16-ch module for thermal neutron detection using ZnS:${}^6$LiF scintillator with embedded WLS fibers coupled to SiPMs and its dedicated readout electronics}

\author{J.-B. Mosset\corref{cor}}
\ead{jean-baptiste.mosset@psi.ch}
\author{A. Stoykov, U. Greuter, A. Gromov, M. Hildebrandt, T. Panzner, N. Schlumpf}
\cortext[cor]{Corresponding author}
\address{Paul Scherrer Institut, CH-5232 Villigen PSI, Switzerland}

\begin{abstract}
A scalable 16-ch thermal neutron detection system has been developed in the framework of the upgrade of a neutron diffractometer. The detector is based on ZnS:${}^6$LiF scintillator with embedded WLS fibers which are read out with SiPMs. In this paper, we present the 16-ch module, the dedicated readout electronics, a direct comparison between the performance of the diffractometer obtained with the current ${}^3$He detector and with the 16-ch detection module, and the channel-to-channel uniformity.
\end{abstract}


\end{frontmatter}


\section{Introduction}

POLDI is a neutron time-of-flight (TOF) diffractometer \cite{poldi} commissioned in 2002 at the Swiss neutron spallation source (SINQ) at the Paul Scherrer Institute (PSI). It was designed to function as a strain scanner for the investigation of residual stress in engineering materials and components. While residual stress measurements are still an important part of the beamtime usage, 30\% of the research at POLDI involves now in-situ testing and this will increase with the recent installation of a tension/compression/biaxial deformation rig which is unique and not available at other beamlines. In order to allow a simultaneous measurement of the axial and transverse strain components during in-situ deformation measurements, an upgrade program of the POLDI beamline has been launched which is aiming for the installation of a second detector oppositely placed to the current ${}^3$He detector with the same geometry and the same level of performance. 

\renewcommand{\arraystretch}{1.1}
\begin{table}[h]
\caption{Main requirements for the new POLDI detector.}
\begin{small}
\begin{center}
\begin{tabular}{ll}
\hline
detector size (width $\times$ height) & 1 $\times$ 0.2 m${}^2$ \\
radius of curvature & 2 m\\
number of channels & 400\\
channel size (width $\times$ height) & 2.5 $\times$ 200 mm${}^2$\\
neutron spectrum & 1 - 5 \AA\\
detection efficiency & $>65$\% at 1.2 \AA\\
time resolution & $<2$~$\muup$s \\
sustainable count rate & $>4$~kHz / channel \\
quiet background rate & $< 0.003$ Hz / channel \\
gamma sensitivity & $<10^{-6}$\\
\hline
\end{tabular}
\end{center}
\label{table:requirements}
\end{small}
\end{table}

Table \ref{table:requirements} gives the requirements for the new detector. Because of the world-wide shortage of ${}^3$He, a simple duplication of the current detector was not possible. As an alternative to the ${}^3$He technology, we decided to use the ZnS/${}^6$LiF scintillator technology.

\section{The 16-ch detection module}

\begin{figure}[h!]
\centering
\includegraphics[width=0.8\linewidth]{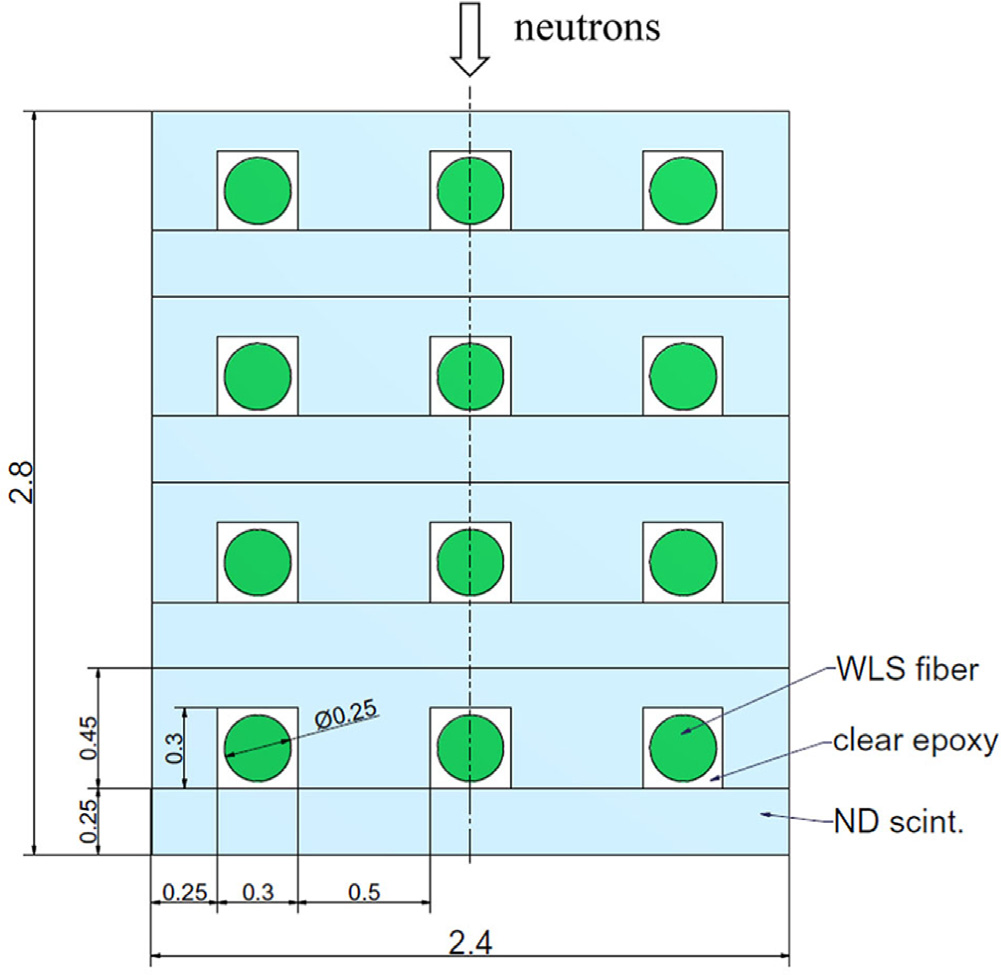}
\includegraphics[width=0.4\linewidth]{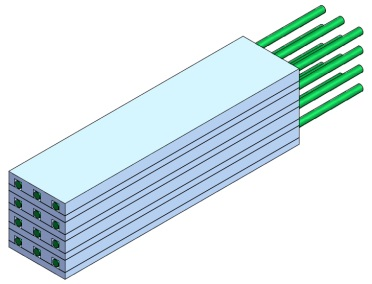}
\caption{The upper figure shows the cross-section of a detection bar consisting of a layered assembly of 0.25~mm and 0.45~mm thick scintillation screens. The optical epoxy EJ-500 from Eljen Technology \cite{eljen} is used to glue toghether the scintillation screens as well as the WLS fibers in the grooves of the 0.45~mm thick scintillation screens. The bottom figure shows a 3 dimensional view of a detection bar. The bar is 2.4~mm wide, 2.8~mm high, and 200~mm long. It represents the elementary building block of a multichannel detector.}
\label{figure:structure}
\end{figure}

The 16-ch module has been described in detail recently in \cite{hildebrandt_pisa}. We repeat here the main aspects. The 16-ch module consists of an almost gapless assembly of 16 individual detection bars with a pitch of 2.5~mm. A detection bar is a layered assembly of scintillation screens ND2:1 from Scintacor \cite{scintacor} with embedded wavelength-shifting (WLS) fibers Y11(400)M from Kuraray \cite{kuraray} uniformly distributed in the detection volume. This structure shown in Fig. \ref{figure:structure} provides an efficient and uniform light collection efficiency despit of the poor light transmission of the scintillator as well as a large neutron absorption probability of 84\% at 1.2~\AA. On one extremity of the bar the surface is polished and mirrored while on the other the 12 WLS fibers are bundled and glued into a 1.1~mm diameter hole of a plexiglas block before they are polished.

\begin{figure}[h!]
\begin{minipage}{0.49\linewidth}
\includegraphics[width=1.\linewidth, angle=180]{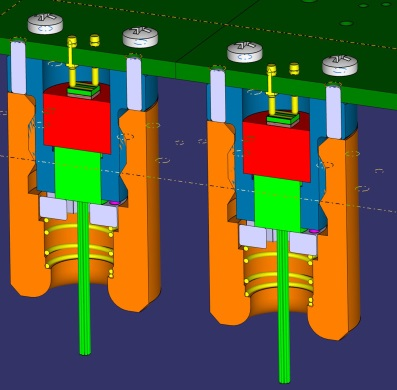}
\caption{Optical coupling between WLS fibers and SiPMs. To ensure a uniform illumination of the SiPMs, each WLS fiber bundle is coupled to a SiPM through a 2~mm long clear multiclad fiber with a diameter of 1.2~mm. The SiPMs are connected to the front-end board with spring contacts and the plexiglas blocks holding the WLS bundles are pressed with a spring. This offers the possibility to exchange easily the SiPMs.}
\label{figure:optical_coupling}
\end{minipage}
\hspace{0.02\linewidth}
\begin{minipage}{0.47\linewidth}
\includegraphics[width=1.\linewidth]{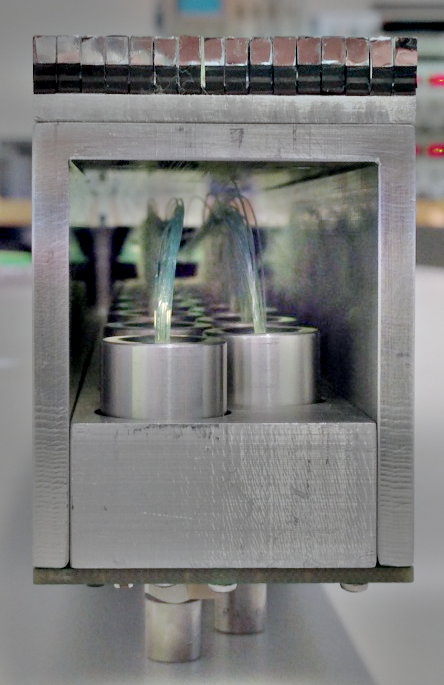}
\caption{Side view of the 16-ch detection module. The figure shows the mirrored extremities of the detection bars and the bundles of 12 WLS fibers running inside the module to the SiPMs.}
\label{figure:16-ch_module}
\end{minipage} 
\end{figure}

The fiber bundle is coupled to a SiPM S12571-025C from Hamamatsu \cite{hamamatsu} through a 2~mm long clear multiclad fiber with a diameter of 1.2~mm which is polished on both ends (see Fig. \ref{figure:optical_coupling}). This optical coupling provides a uniform illumination of the $1 \times 1$~mm${}^2$ sensitive area of the SiPM independent of which WLS fibers collect the light. The SiPMs are connected with spring contacts to the front-end board. This offers the possibility to exchange the SiPMs in case of a necessary replacement due to radiation damage or the use of new devices with improved properties. Fig. \ref{figure:16-ch_module} shows the 16-ch module. Each detection bar is mounted individually and can be replaced in case of damage without affecting the others. To prevent optical crosstalk between adjacent bars, each one is wrapped with a thin aluminum foil.

The front-end board to which the SiPMs are connected contains a passive thermistor circuit \cite{thermistor_circuit} on each high voltage line to compensate for temperature variations. This circuit stabilizes the SiPM overvoltage over a large range of temperature from about 15\textdegree C to 45\textdegree C.

\section{Signal processing}

The signal processing system is based on a photon counting approach. Two reasons make this approach feasible. First, the ZnS:${}^6$LiF scintillator is relatively slow \cite{kuzmin} and consequently the photons are sufficiently spaced in time to allow the detection of almost each individual photon emitted in a given time window. Secondly, SiPMs have an excellent single photon counting capability. 

\begin{figure}[h!]
\centering
\includegraphics[width=1.0\linewidth]{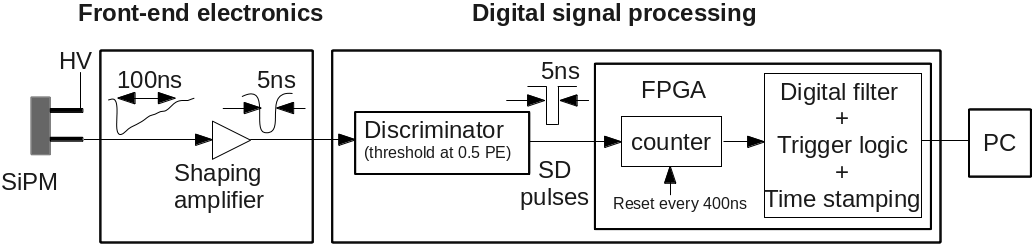}
\caption{Block diagram of the signal processing system. A detailed description is given in the text.}
\label{figure:SPS}
\end{figure}

\begin{figure}[h!]
\centering
\includegraphics[width=0.6\linewidth]{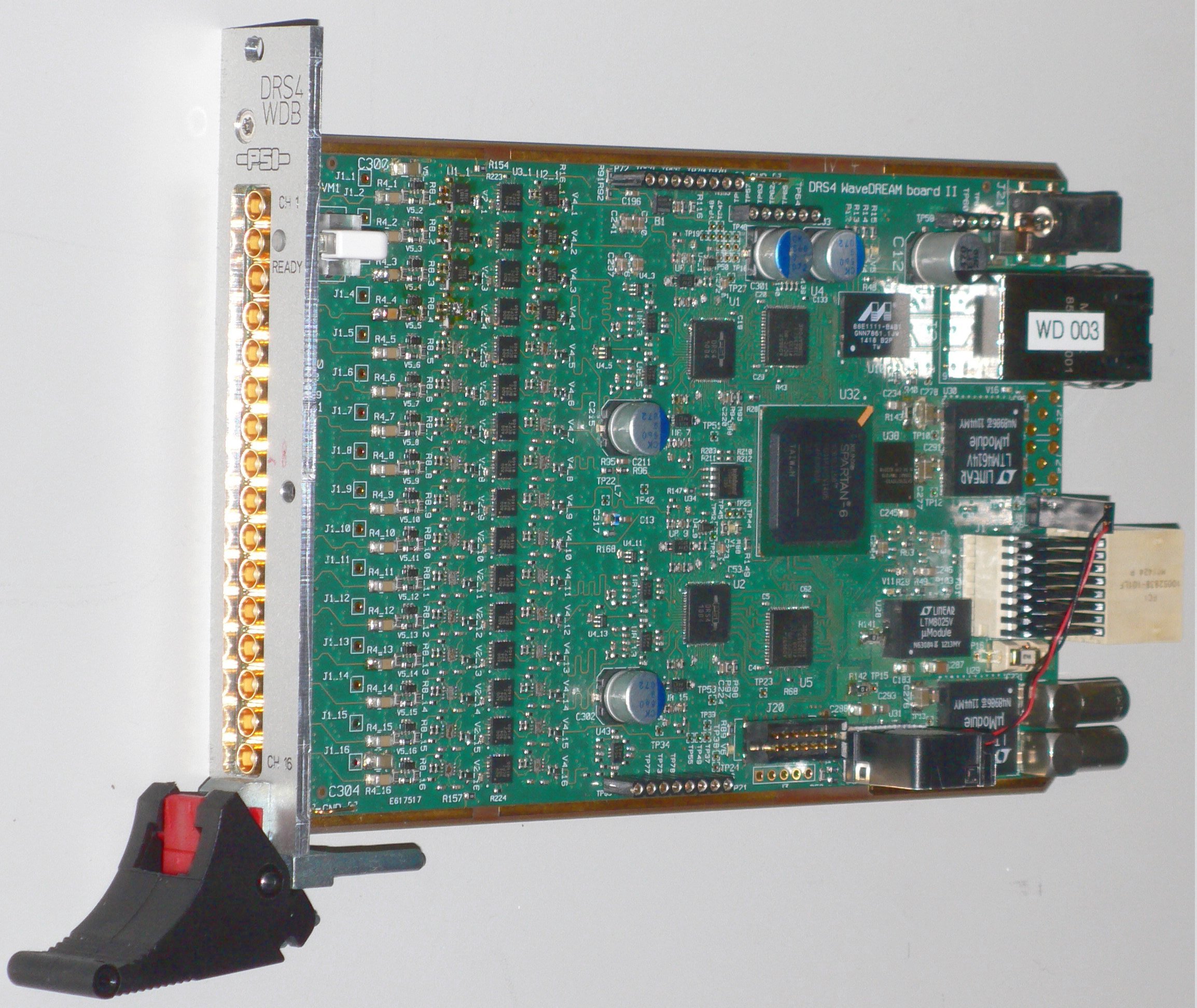}
\caption{The WaveDREAM board: a 16-ch FPGA based readout electronics.}
\label{figure:wavedream}
\end{figure}

Fig. \ref{figure:SPS} shows the block diagram of the signal processing system. The SiPM output is connected to a high frequency amplifier producing a 5~ns wide pulse for each single-cell signal of the SiPM. The output of the amplifier is then fed into a leading edge discriminator with a threshold set at 0.5 photoelectron (PE). Independent of how many SiPM cells are triggered simultaneously due to the SiPM crosstalk, the discriminator creates only one standard SD-pulse. This ``digitization'' which suppresses the crosstalk allows to reduce the quiet background rate of the detector \cite{mosset764}.

The analog output signals from the front-end board are readout by the WaveDREAM board, a 16-ch FPGA-based readout board (Fig. \ref{figure:wavedream}) developed at PSI for the MEG experiment. The WaveDREAM board is highly flexible and offers all the necessary features for this project: discriminators and FPGA. The board is connected to a PC via a GBit Ethernet connection which offers a fast data transfer. We achieve a maximum count rate of neutrons equal to 4~kHz/ch without loss, which satisfies the requirements for POLDI.

The nature of the signal received by the digital filter is the temporal SD-pulse density. This signal is sampled every 400~ns by measuring the number of SD-pulses during each consecutive time slice of 400~ns. In \cite{mosset_pisa}, the performances of the detector obtained with several digital filters are compared. The best evaluated filter, the so-called ``moving sum after differentiation'' (DI+MS) filter, is the one which has been implemented in the WaveDREAM board. It is defined by the following formulas:\vspace{1.mm}
\begin{equation}
z_i = z_{i-1} + y_{i} - y_{i-M} \hspace{10mm} y_i = x_{i} - x_{i-M} 
\end{equation}
where $x_i$ and $z_i$ are the discrete samples of the signal at the input and output of the filter and $M$ is a parameter which determines the shaping time of the filter equal to $M \times 400$~ns.

After calculation of the filter output, the following triggering conditions are ANDed: (a)~The channel is ready (after each trigger, a blocking time is introduced to prevent multiple triggers on the same events). (b)~The filter output is maximum. (c)~The filter output is higher than a certain threshold. 

The time stamp of an event is defined by the time slice in which the filter output is maximum.

The detection module is operated with the following settings of the signal processing: $M$=5, blocking time = 4~$\mu$s, threshold = 20~SD-pulses. With this settings and up to a SiPM dark count rate of 4~MHz, the performance parameters of the detection module reported in \cite{mosset_pisa} are: trigger efficiency = 81\%, background rate $<10^{-3}$~Hz/ch, probability of multiple trigger $<10^{-3}$, time resolution (StDev) $<170$~ns (contribution from the signal processing). Based on recent measurements performed with an analog CR-RC${}^4$ filter, which is equivalent to the DI+MS filter, we estimate the gamma sensitivity of the detection module to $10^{-7}$ up to a SiPM dark count rate of 4~MHz. 

The initial dark count rate of the SiPMs is 100~kHz and irradiation tests in the POLDI beamline show an increase of $\sim$90~kHz/year of the SiPM dark count rate. So, the detection module could be operated during more than 20~years with the above-mentioned performance.

\section{Measurements}

\subsection{Diffractometer performance with current and new detector}

In order to obtain a direct comparison between the current ${}^3$He detector and the new 16-ch detection module, the latter was placed oppositely to the current ${}^3$He detector, at a scattering angle of about {90\textdegree} and at the same distance from the sample as the ${}^3$He detector. Both detectors were synchronized so that they started and stopped their acquisition of data simultaneously and their timers were reset simultaneously at the beginning of each chopper sequence.

There was no available collimator for the 16-ch module. Therefore, in order to have identical measurement conditions for both detectors, the collimator of the ${}^3$He detector was removed and a small size sample which does not require a collimator was used (Iron wire of 1~mm diameter) .

\begin{figure}[h!]
\centering
\includegraphics[width=1.0\linewidth]{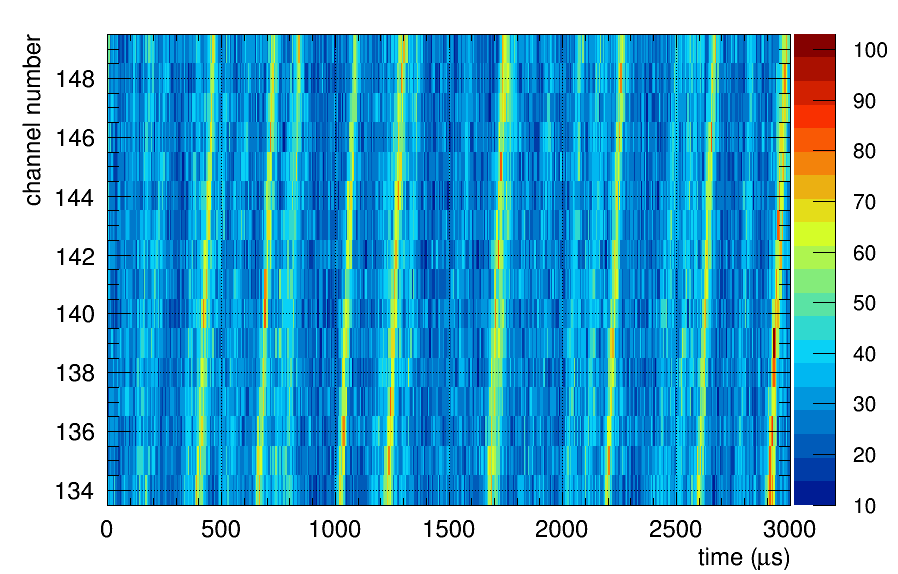}
\includegraphics[width=1.0\linewidth]{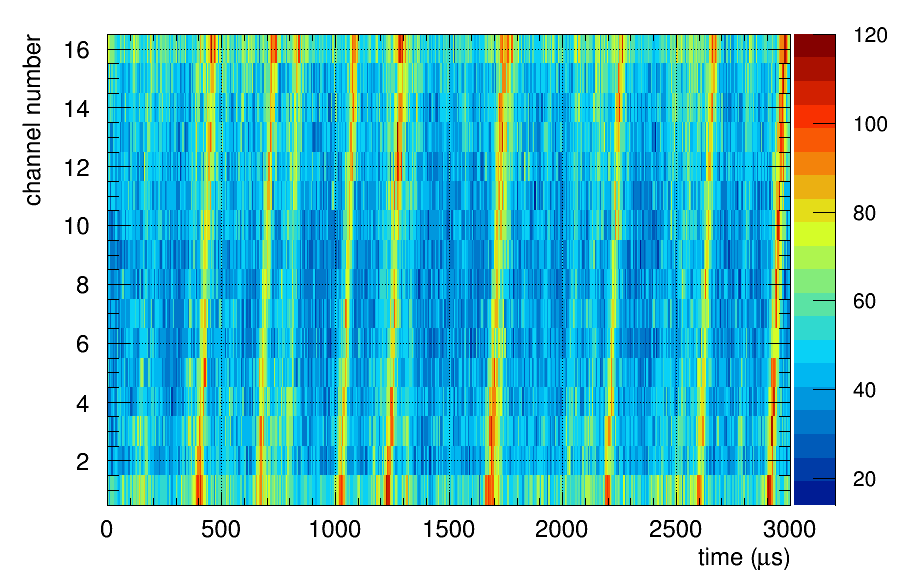}
\caption{Raw data obtained with the current ${}^3$He detector (top) and with the 16-ch scintillation detector (bottom). The time scale represents the time elapsed since the last reset of the timers done by an index pulse which is sent to the detectors at the beginning of a chopper sequency. A chopper sequency contains 8 slits with a non-periodic distribution. Hence, each Bragg peak results into 8 lines.}
\label{figure:raw_data}
\end{figure}

In order to have a fair comparison of the diffractometer performance obtained with the ${}^3$He and with the scintillation detectors, we only used the 16 channels of the ${}^3$He detector (channels 134 to 149) having the same angular positions as the ones of the scintillation detector. Fig. \ref{figure:raw_data} shows the raw data obtained with the two detectors.

With the scintillation detector, the number of events located between the Bragg lines (background) appears to be higher than with the ${}^3$He detector, in particular for the channels located on the edge of the detector (channels 1 and 16). The reason for that is not so much related to the detection units themselves but mainly to their shielding against neutrons scattered from somewhere else than the sample, or against fast neutrons. The ${}^3$He detector had a better shielding than the scintillation detector. Moreover, in the ${}^3$He detector, the detection channels surrounding the selected channels were acting as a shielding providing a uniform background, in contrast to the situation with the 16-ch scintillation detector.

\begin{figure}[h!]
\centering
\includegraphics[width=1.0\linewidth]{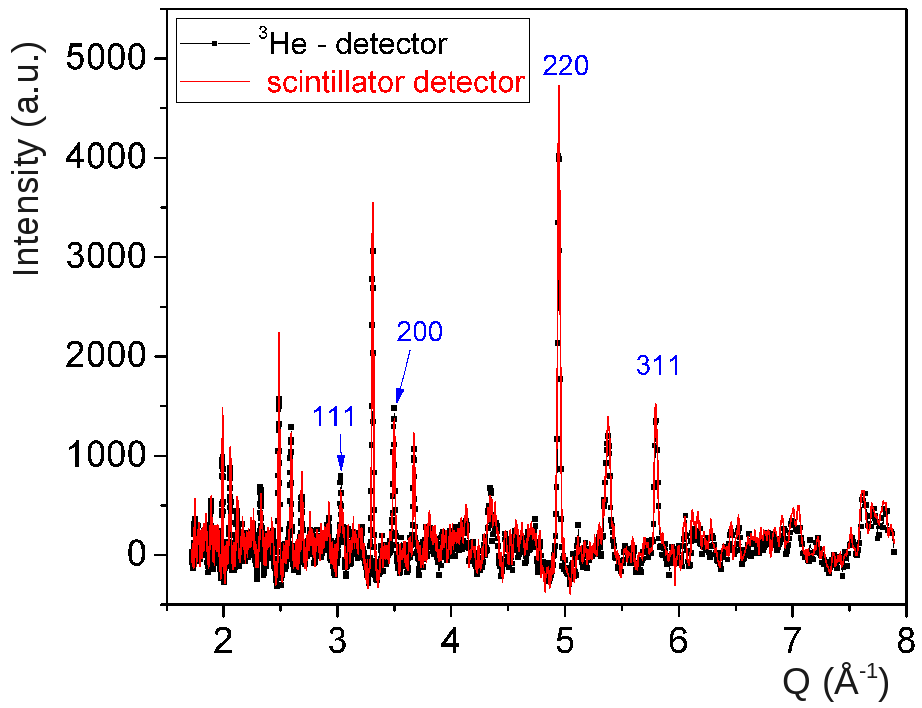}
\caption{Correlation diffraction patterns obtained with the 16-ch scintillation detector and with the 16 corresponding channels of the ${}^3$He detector.}
\label{figure:diffraction_pattern}
\end{figure}

Fig. \ref{figure:diffraction_pattern} shows the correlated diffraction pattern obtained with the 16-ch scintillation detector and with the 16 corresponding channels of the ${}^3$He detector. Only four peaks correspond to Bragg peaks. The others are artifacts created by the algorithm due to the limited number of channels used. Nevertheless, we observe a perfect agreements concerning the peak positions and the diffractometer resolution (FWHM of the Bragg peak divided by its position) obtained with the scintillation detector is equivalent or even better than with the current ${}^3$He detector.

\subsection{Channel-to-channel uniformity}

\begin{figure}[h!]
\centering
\includegraphics[width=1.0\linewidth]{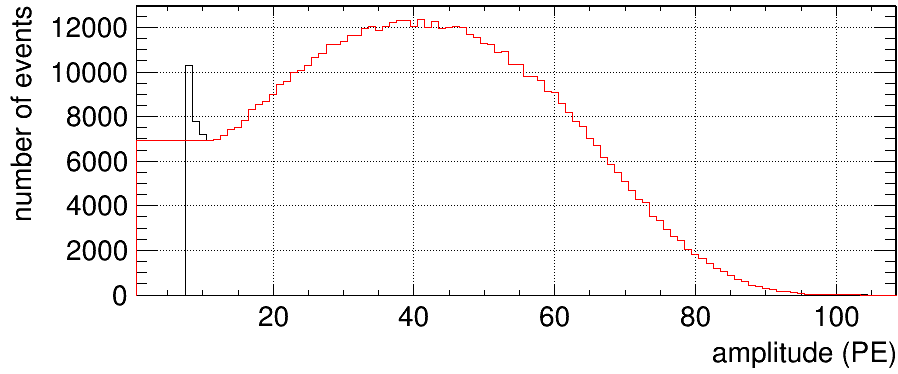}
\caption{The black curve shows the amplitude spectrum for one of the 16-ch prototype measured at a low trigger threshold. At low amplitude, the spectrum is a little bit contaminated with background events due to the dark count rate of the SiPM. The red curve corresponds to the measured amplitude spectrum down to the position of the first minimum of the spectrum ($\sim$11~PE) where the background contamination is smaller. From this amplitude ($\sim$11~PE) down to 0~PE, the amplitude spectrum is extrapolated with a plateau at the level of the first minimum.}
\label{figure:spectrum}
\end{figure}

Fig. \ref{figure:spectrum} shows the amplitude spectrum for one channel of the 16-ch module measured at a low trigger threshold. The trigger efficiency at a threshold $A_{threshold}$ is calculated with the equation
\begin{equation}
\epsilon_{trigger} = 1 - \frac{\int_{0}^{A_{threshold}} h}{N_{events}} \label{eq:trigger_efficiency} 
\end{equation}
where $h$ is the extrapolated amplitude spectrum and $N_{events}$ is the total number of events in $h$.

\begin{figure}[h!]
\centering
\includegraphics[width=1.0\linewidth]{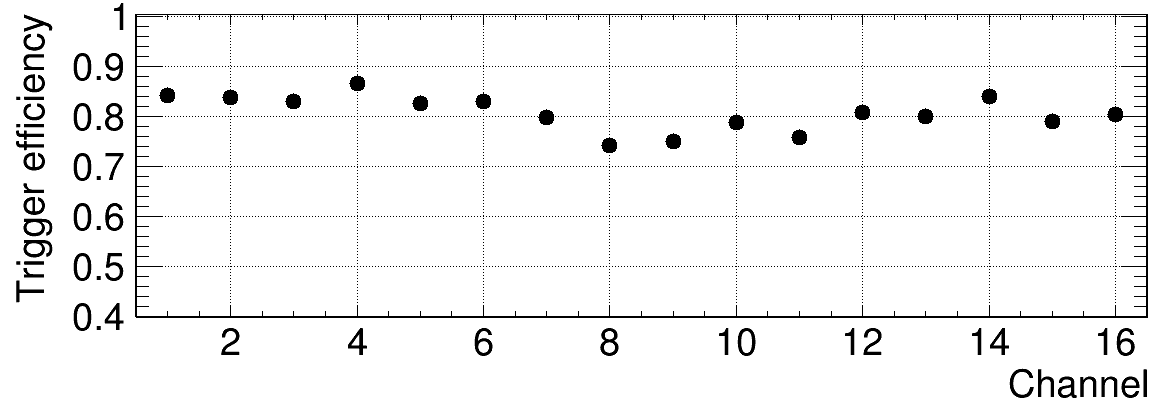}
\caption{Tigger efficiency profile of the 16-ch detection module.}
\label{figure:uniformity}
\end{figure}

Fig. \ref{figure:uniformity} shows the tigger efficiency profile of the 16-ch detection module. The average and the standard deviation of the trigger efficiencies of the individual channels are 81\% and 3.4\%, respectively.

\section{Summary}

The upgrade program of the POLDI engineering diffractometer at PSI is aiming to the installation of an additional detector with the same geometry and the same level of performance as the current ${}^3$He detector. In order to tackle the problem of the ${}^3$He shortage, we developped a 16-ch detection module based on ZnS(Ag):${}^6$LiF scintillator with WLS fibers as well as its dedicated digital signal processing system based on a photon counting approach. This module represents a building block of the full-size detector. It presents two innovative features for a ZnS(Ag):${}^6$LiF scintillator-based detector that are the layered assembly of the sensitive volume with embedded WLS fibers and the use of SiPMs. The relative standard deviation of the trigger efficiency over the 16 channels is 4\%. This good channel-to-channel uniformity validates the manufacturing process. The resolution of the POLDI diffractometer obtained with the 16-ch scintillation module is equivalent or even better than with the current ${}^3$He detector.

\section*{Acknowledgements}

The authors wish to thank Dieter Fahrni and Andreas Hofer for the design and construction of the neutron detection module. This work was supported by the Swiss National Science Foundation (Grant no. 206021-139106).

\bibliography{mybibfile_VCI_2016_proceeding_Mosset}

\begin{thebibliography}{10}
\expandafter\ifx\csname url\endcsname\relax
  \def\url#1{\texttt{#1}}\fi
\expandafter\ifx\csname urlprefix\endcsname\relax\def\urlprefix{URL }\fi
\expandafter\ifx\csname href\endcsname\relax
  \def\href#1#2{#2} \def\path#1{#1}\fi

\bibitem{poldi}
{U. Stuhr, et al.}, {Time-of-flight diffraction with multiple frame overlap
  Part II: The strain scanner POLDI at PSI}, Nuclear Instruments and Methods in
  Physics Research A 545 (2005) 330.

\bibitem{hildebrandt_pisa}
{M. Hildebrandt, et al.}, {Detection of thermal neutrons using ZnS:Ag/${}^6$LiF
  neutron scintillator read out with WLS fibres and SiPMs}, Nuclear Instruments
  and Methods in Physics Research A (2015),
  http://dx.doi.org/10.1016/j.nima.2015.10.102.

\bibitem{scintacor}
http://www.scintacor.com.

\bibitem{kuraray}
http://kuraraypsf.jp.

\bibitem{eljen}
http://www.eljentechnology.com.

\bibitem{hamamatsu}
http://www.hamamatsu.com.

\bibitem{thermistor_circuit}
{H. Miyamoto, et al.}, {SiPM development and application for astroparticle
  physics experiments}, {Proceedings of the 31\textsuperscript{st} ICRC, 2009}.

\bibitem{kuzmin}
{E.S. Kuzmin, et al.}, {Detector for the FSD Fourie-diffractometer based on
  ZnS(Ag)/${}^6$LiF scintillation screen and wavelength shifting fiber
  readout}, Journal of Neutron Research 10~(1) (2002) 31.

\bibitem{mosset764}
{J.-B. Mosset, et al.}, {Evaluation of two thermal neutron detection units
  consisting of ZnS/${}^6$LiF scintillating layers with embedded WLS fibers
  readout with a SiPM}, Nuclear Instruments and Methods in Physics Research A
  764 (2014) 299.

\bibitem{mosset_pisa}
{J.-B. Mosset, et al.}, {Digital signal processing for a thermal neutron
  detector using ZnS(Ag):${}^6$LiF scintillating layers read out with WLS
  fibers and SiPMs}, Nuclear Instruments and Methods in Physics Research A
  (2015), http://dx.doi.org/10.1016/j.nima.2015.11.062.

\end{thebibliography}
\end{document}